\documentclass[runningheads]{llncs}
\usepackage[utf8]{inputenc}
\usepackage{graphicx}
\usepackage{amsmath}
\usepackage{hyperref}
\usepackage{cleveref}
\usepackage{float}
\usepackage{csquotes}
\begin{document}
\title{Event detection in Colombian security Twitter news using fine-grained latent topic analysis}
\titlerunning{Event detection in Twitter news}
%
\author{Vladimir Vargas-Calderón\inst{1}\orcidID{0000-0001-5476-3300} \and
Nicolás~Parra-A.\inst{1}\orcidID{0000-0002-1829-4399} \and
Jorge E. Camargo\inst{2}\orcidID{0000-0002-3562-4441} \and Herbert~Vinck-Posada\inst{1}}
\authorrunning{V. Vargas-Calderón et al.}
%
\institute{Grupo de Superconductividad y Nanotecnología, Departamento de Física, Universidad Nacional de Colombia, 111321, Colombia\\ \email{\{vvargasc,nparraa,hvinckp\}@unal.edu.co}\and
UnSecureLab Research Group, Departamento de Ingeniería de Sistemas e Industrial, Universidad Nacional de Colombia, 111321, Colombia\\
\email{jecamargom@unal.edu.co}}
\maketitle              
\begin{abstract}
Cultural and social dynamics are important concepts that must be understood in order to grasp what a community cares about. To that end, an excellent source of information on what occurs in a community is the news, especially in recent years, when mass media giants use social networks to communicate and interact with their audience. In this work, we use a method to discover latent topics in tweets from Colombian Twitter news accounts in order to identify the most prominent events in the country. We pay particular attention to security, violence and crime-related tweets because of the violent environment that surrounds Colombian society. The latent topic discovery method that we use builds vector representations of the tweets by using FastText and  finds clusters of tweets through the K-means clustering algorithm. The number of clusters is found by measuring the $C_V$ coherence for a range of number of topics of the Latent Dirichlet Allocation (LDA) model. We finally use Uniform Manifold Approximation and Projection (UMAP) for dimensionality reduction to visualise the tweets vectors. Once the clusters related to security, violence and crime are identified, we proceed to apply the same method within each cluster to perform a fine-grained analysis in which specific events mentioned in the news are grouped together. Our method is able to discover event-specific sets of news, which is the baseline to perform an extensive analysis of how people engage in Twitter threads on the different types of news, with an emphasis on security, violence and crime-related tweets.

\end{abstract}

\section{Introduction}

Twitter\footnote{\url{http://twitter.com}} is a social network that has been used worldwide as a means of news spreading. In fact, more than 85\% of its users use Twitter to be updated with news, and do so on a daily basis~\cite{rosenstiel2015twitter}. Users behaviour of this social network has been found efficient in electronic word-of-mouth processes~\cite{williams2015community}, which is a key component for the quick spreading of breaking news. This would lead to think that news-related content occupies the majority of the tweets volume. However, on average, the proportion of news-related content to the total content of tweets is 1\% worldwide, but increases dramatically (up to 15\%) in countries in conflict~\cite{malik2016}. An extrapolation of these findings indicates that Colombia might have a high content of news-related tweets, since it is well-known that Colombia is one of the most violent countries in the world, and has been for decades~\cite{index2018measuring}.

On the other hand, the virality or importance of a tweet conveying news-related information is a relevant measure of what is critical for a community. Therefore, the study of news spreading in a community gives a clear idea of citizens interactions around central topics of interest. Particularly, we are interested in examining how people react to news related to security, crime and violence because this would expose the mechanisms of collective reactions of rejection, acceptance, conflict, among others. This has been considered in case-studies such as Ref.~\cite{miro2016cyber}, where messages containing hate or violent speech were identified after Charlie Hebdo's famous shooting, allowing researchers to identify spatio-temporal and textual patterns in the produced tweets after the mentioned disruptive event. Other similar case-studies include the analysis of how people react to homicides in London~\cite{kounadi2015exploring}, to polio health news~\cite{schaible2017twitter}, and to the aftermath of violence on college campuses~\cite{jones2016tweeting}. Also, social networks and technology have been signalled as tools used by young people to inflict violent acts against others~\cite{draucker2010role,julia2013,mengu2015violence}. On a more general ground, the study of these individual or collective reactions is a problem tackled by sentiment analysis, whose objective is to determine whether the sentiment contained in a text is positive or negative, and to what extent~\cite{rosenthal2017semeval,agarwal2011sentiment,nakov2016semeval,pak2010twitter}. Applied to security-related content in social networks, sentiment analysis could be important in designing and implementing public policies regarding security, crime and violence, as well as educational campaigns where people are taught to communicate their opinions in a non-violent way. However, in order to achieve this, it is desirable to segment news-related tweets so that different topics can be differentiated from one another as we expect Twitter users to react quite differently depending on the security topic they react to. This field is known as topic discovery.

Several proposals for topic discovery are available, among which many Latent Dirichlet Allocation variants and modifications are available. For instance, Ref. \cite{zhou2017} presents an LDA-based model that relates the topic of a scientific paper with the content of the documents that it cites. The proposed algorithm allows to know the evolution of a research topic by measuring whether a topic is important (as seen by the scientific community) or not. Furthermore, Ref. \cite{groof2017} used LDA with variational Gibbs sampling to found general terms that associate  the reviews of users in e-commerce web sites. This with the intention of improving the experience of new users. Moreover, we have recently combined word-embedding methods and K-means to discover topics and have good interpretability results \cite{vargascaldern2019using,vargas2019characterization}.

Thus, in this paper we exemplify a method for topic discovery applied to Colombian news-related tweets that is accurate in the task of segmenting tweets. This method can work at different granularity levels depending on the corpus to be analysed. In this case, we have a corpus of security-related tweets, so that the method will group tweets into the different sub-topics such as murders, robberies, among others. The workings of the method will be detailed in section~\cref{sec:methods}, and the main results will be presented in Section~\cref{sec:results}. Finally, we provide some conclusions in Section~\cref{sec:conclusions}.

\section{Methods and Materials\label{sec:methods}}

In this section, we describe the dataset used in our research, as well as the methods to perform fine-grained latent topic analysis to process all the data. The method is largely based on our previous work~\cite{vargascaldern2019using}.

Tweets from Colombian news Twitter account @NoticiasRCN were collected from 2014 to the present. A total of 258,848 tweets were published by @NoticiasRCN in this period. The method hereafter mentioned was applied in Ref.~\cite{vargascaldern2019using} to this large corpus at a coarse-grained scale to discover news topics, allowing us to pinpoint groups of tweets sharing semantic content. It was possible to detect tweets regarding to politics, sports, Colombian armed conflict, extreme violence, organised/common crime, among others. In this paper, we focus on the groups of extreme violence and organised/common crime, which contain a total of 47,229 tweets, accounting for an 18.2\% of all published news. We excluded the Colombian armed conflict, as this is not normally connected to events occurring in cities.

We pre-processed these tweets related to security, crime and violence by removing punctuation, links, hashtags and mentions, we lowercased the text and performed lemmatisation with spaCy's adapted Spanish lemmatiser\footnote{\url{https://github.com/lang-ai/es_lemmatizer}}.

As the objective of our work is to find a number of topics and its members that are helpful for further analysis, the first task to solve is to methodologically find the number of topics. Our proposal is to combine a topic modelling tool with a measure of how well this tool performed. Therefore, we trained a Latent Dirichlet Allocation (LDA) model~\cite{blei2003latent} to soft-cluster tweets into topics and then used $C_V$ coherence~\cite{roder2015exploring,syed2017full} to measure the interpretability of the LDA results. What LDA does is to assign to documents probabilities to belong to different topics (an integer number $k$ provided by the user), where these probabilities depend on the occurrence of words which are assumed to co-occur in documents belonging to the same topic. This assumption is called a sparse Dirichlet prior. Thus, LDA exploits the fact that even if a word belongs to many topics, occurring in them with different probabilities, they co-occur with neighbouring words in each topic with other probabilities that help to better define the topics. The best number of topics is the number of topics that helps the most human interpretability of the topics. This means that if the topics given by LDA can be well-distinguished by humans, then the corresponding number of topics is acceptable. As mentioned before, a way of measuring this interpretability is the calculation of $C_V$ coherence, which, to the knowledge of the authors, is the measure with largest correlation to human interpretability. The optimum number of topics can be found at the maximum of $C_V(k)$.

Once the number of topics has been determined, we proceed to find vector-embedding representations of tweets, as they have been previously shown to yield superior results in topic modelling with respect to LDA~\cite{he2018vector}. Here, we use the word2vec-based~\cite{mikolov2013distributed,goldberg2014word2vec,rong2014word2vec} FastText model~\cite{bojanowski2016}, which essentially uses sub-word information to enrich embeddings generated by a neural network that predicts neighbouring words. The job of FastText is to reduce the dimensionality of one-hot-encoded words (which may be in very large vector spaces of the size of the corpus vocabulary) to a low-dimension and dense vector space (of dimension $N$, selected by the user), where dimensions store highly correlated semantic relations between words and strings of characters. Here a tweet is represented by the sum of the individual vector representations of each word in the tweet.

In this low-dimensional vector space, K-means clustering~\cite{lloyd1982} is performed for $k$ clusters (i.e. the number of topics found with the LDA-$C_V$ coherence method) in order to hard-cluster the vector representations of the tweets. K-means is a common clustering technique that minimises within-cluster dispersion. Each cluster contains tweets belonging to the same topic.

In order to visualise the clusters and interpret their contents, we performed dimensionality reduction with the Uniform Manifold Approximation and Projection (UMAP) method~\cite{mcinnes2018umap} to plot vectors onto a 2 dimensional space. UMAP learns the topology of the data to be reduced in dimension by learning a projection of this data onto a lower-dimensional space where the projection preserves, as much as possible, the fuzzy topological structure of the manifold described by the vectors. Additionally, the Python's Bokeh library~\cite{bokeh2018} was used to create interactive plots of the UMAP reduced representation of FastText vectors, allowing us to quickly examine the structure of the clusters, as well as to read representative tweets of each cluster to determine and label their corresponding topics.

\section{Results and Discussion\label{sec:results}}

\begin{figure}[!ht]
    \centering
    \includegraphics[width=\textwidth]{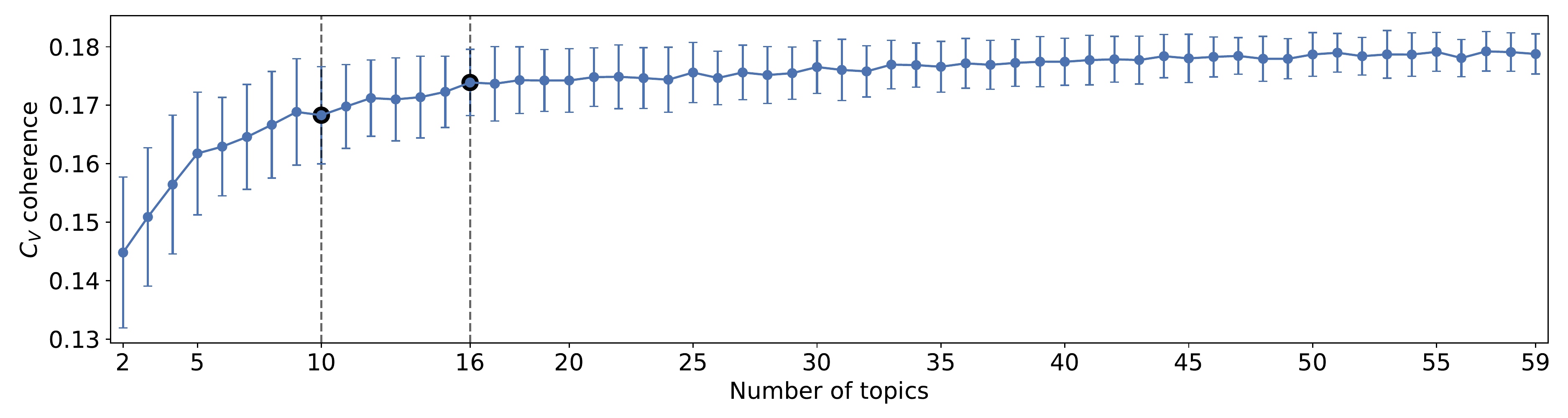}
    \caption{$C_V$ coherence as a function of the number of topics. Error bars indicate one standard deviation error.}
    \label{fig:cv}
\end{figure}

Since LDA is a probabilistic method that starts its learning with some random parameters, we measured $C_V$ coherence 64 times for each number of topics $k$ ranging from 2 to 59, and averaged the measurements. The results are shown in~\cref{fig:cv}. It is important to note that a maximum is not reached. However, a saturation of the $C_V$ coherence takes place, making it difficult to select a number of topics. By visual inspection and the use of the so-called elbow method, we pick two different numbers of topics (10 and 16) and analyse them separately in order to determine what the best number of topics is.

\begin{figure}[!ht]
    \centering
    \includegraphics[width=\textwidth]{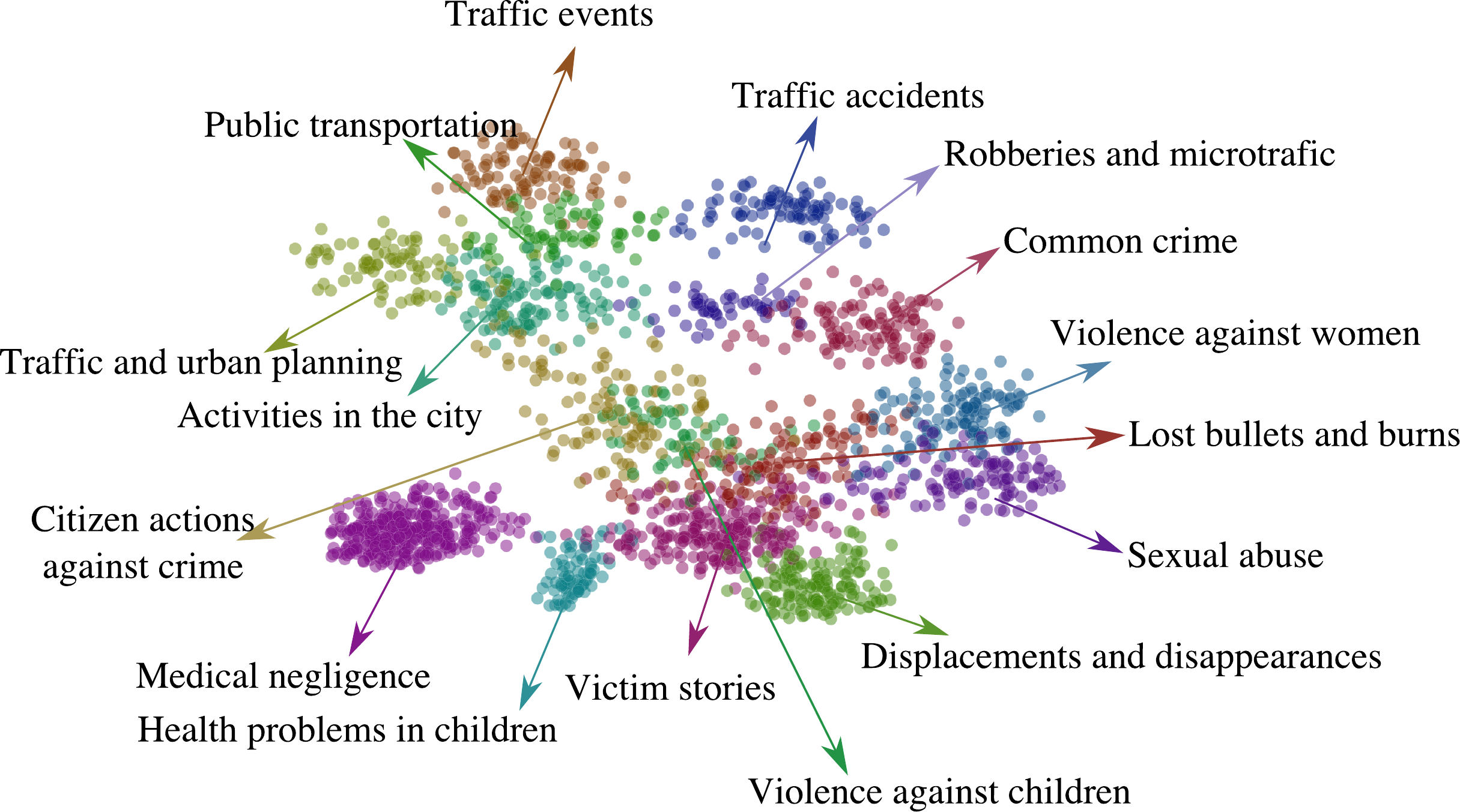}
    \caption{K-means clustering results with 16 topics and their respective names.}
    \label{fig:cluste_16}
\end{figure}

Then, the FastText method was trained using a 30-dimensional vector embedding space. In \cref{fig:cluste_16} we plot the UMAP-reduced vectors for 16 clusters that are at a maximum Euclidean distance of 0.2 (arbitrary units) from their respective cluster centroid, found through K-means clustering. We manually labelled the clusters by reading the 15 most representative tweets of each cluster, i.e. the ones closest to the corresponding cluster centroid. Some clusters were difficult to label, particularly those that clearly overlap with other clusters in the visualisation.

\begin{figure}[!ht]
    \centering
    \includegraphics[width=\textwidth]{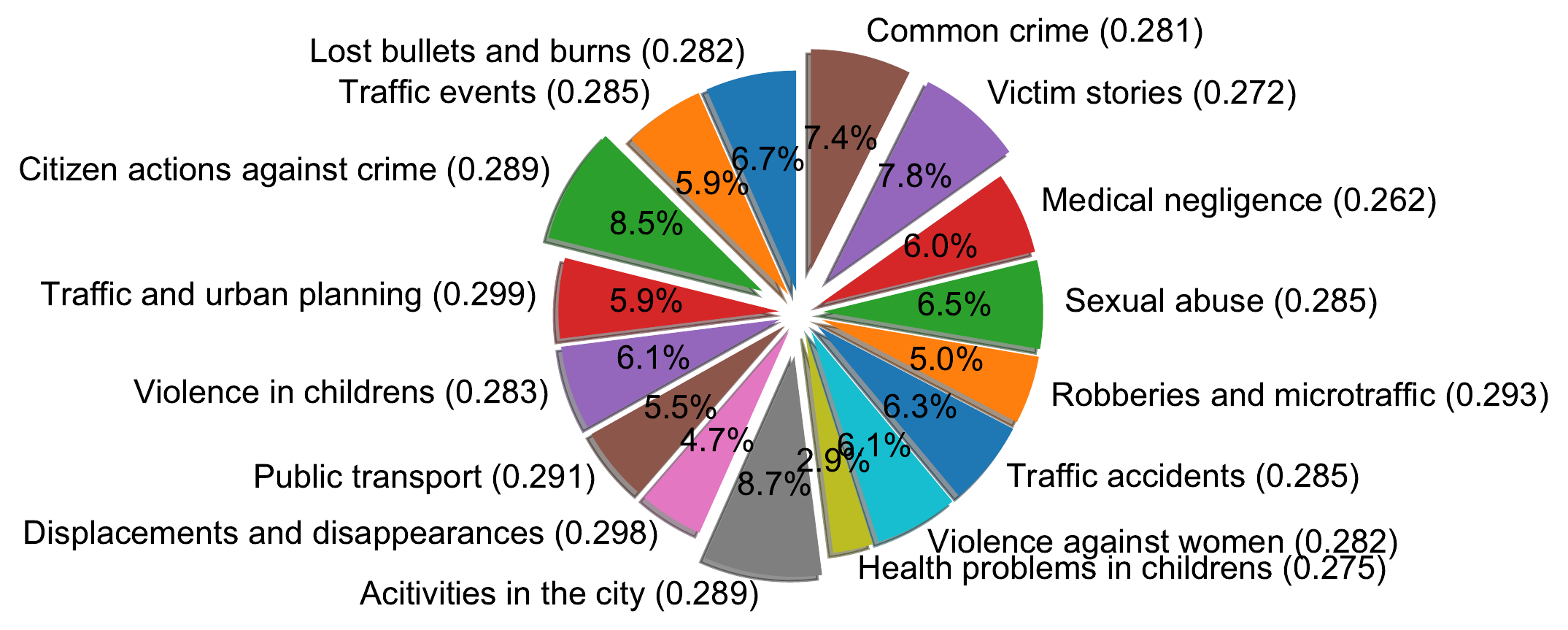}
    \caption{Number of tweet distribution in the 16 topics, the number in parenthesis that follows each label is the within-cluster dispersion $N_{T_i}^{-1}\sum_{k\in T_i}d_{k}^{i}$, where $T_i$ is the set of tweets corresponding to topic $i$, $N_{T_i}$ is the number of tweets for that topic and $d_k^i$ is the distance of the $k$-th tweet vector representation to the centroid of the $i$-th topic.}
    \label{fig:pie_sixteen}
\end{figure}

The distribution of tweets along the 16 topics is shown in~\cref{fig:pie_sixteen}, where news were slightly concentrated on security-related sub-topics like activities in the city, citizen actions against crime, victim stories and common crime. The within-cluster dispersion is comparable between topics, implying that clusters are equally diffused.

\begin{figure}[!ht]
    \centering
    \includegraphics[width=\textwidth]{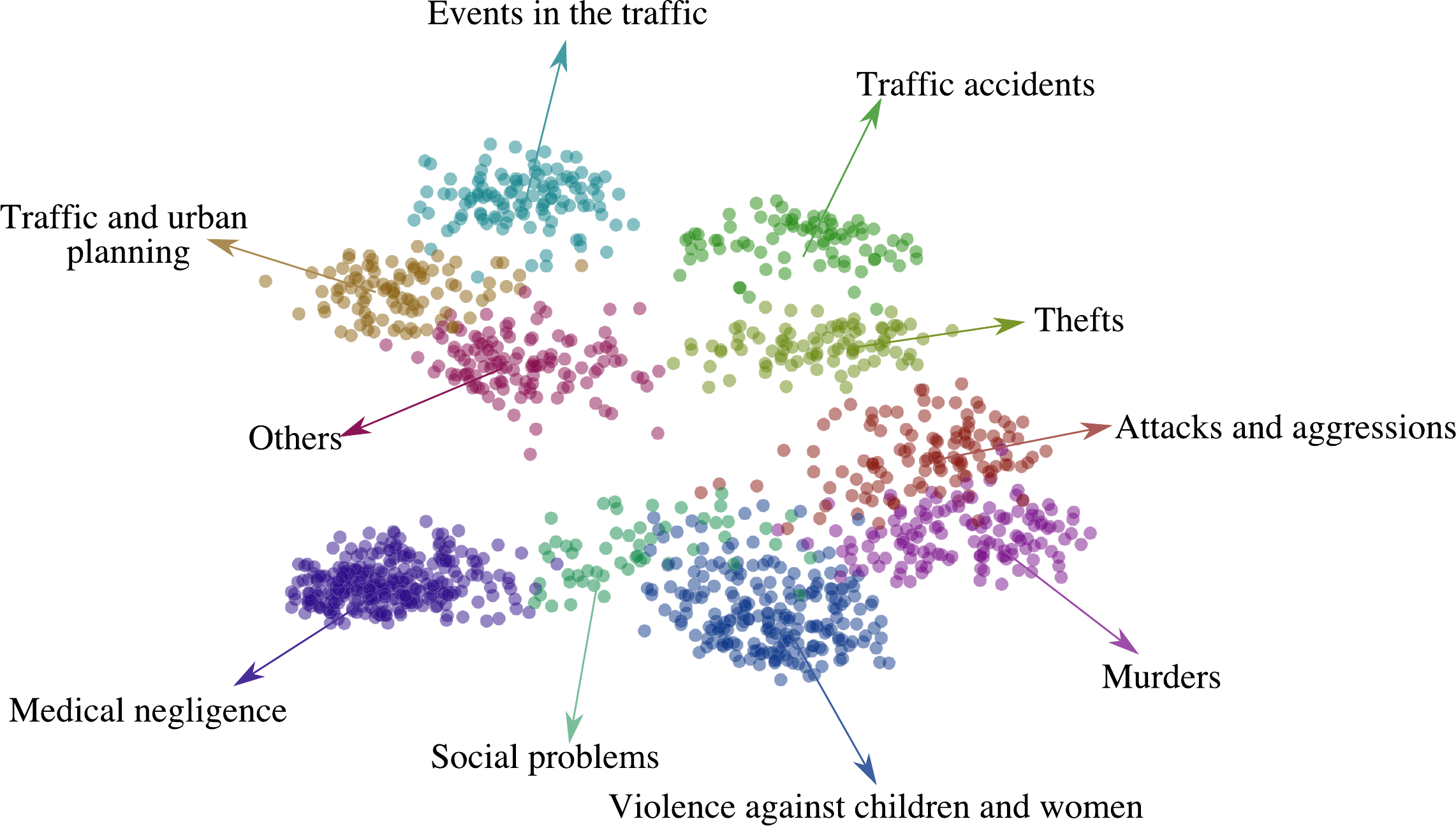}
    \caption{Same as~\cref{fig:cluste_16} but for 10 topics.}
    \label{fig:ten_clusters}
\end{figure}

Regarding the case of only 10 clusters, in~\cref{fig:ten_clusters} we plot the UMAP-reduced vectors that are most representative of each cluster, just as in~\cref{fig:cluste_16}. We find that this number of topics is better than 16 clusters, since the identification of the topic was clearer in the case of 10 clusters when reading the 15 most representative tweets of each cluster. This holds true even for three different topics referring to traffic, as they topics can be well-differentiated (note that UMAP plots those groups close one another because they all contain traffic-related tweets). For instance in the traffic and urban planning cluster we find news such as
\begin{quote}
Bogotá's Secretaría de Movilidad talks about changes that users will find in public transportation this Monday
\end{quote}
which is not directly related with security, crime or violence topics. On the other hand, in the traffic accidents cluster we find news such as
\begin{quote}
In Cartagena, two bus drivers left their vehicles in the middle of the highway to solve their differences by fighting against each other.
\end{quote}
Finally, in the events in the traffic cluster, an example of a news tweet is
\begin{quote}
North highway is collapsed by a triple-crash. The air patroller recommends to take alternate routes.
\end{quote}
These tweets exemplify the difference of the three traffic-related clusters. Moreover, by comparing ~\cref{fig:cluste_16,fig:ten_clusters}, it is clear that some clusters are barely changed when increasing the number of topics from 10 to 16 because they are well-defined topics that can be interpreted easily from the human perspective.

\begin{figure}[!ht]
    \centering
    \includegraphics[width=\textwidth]{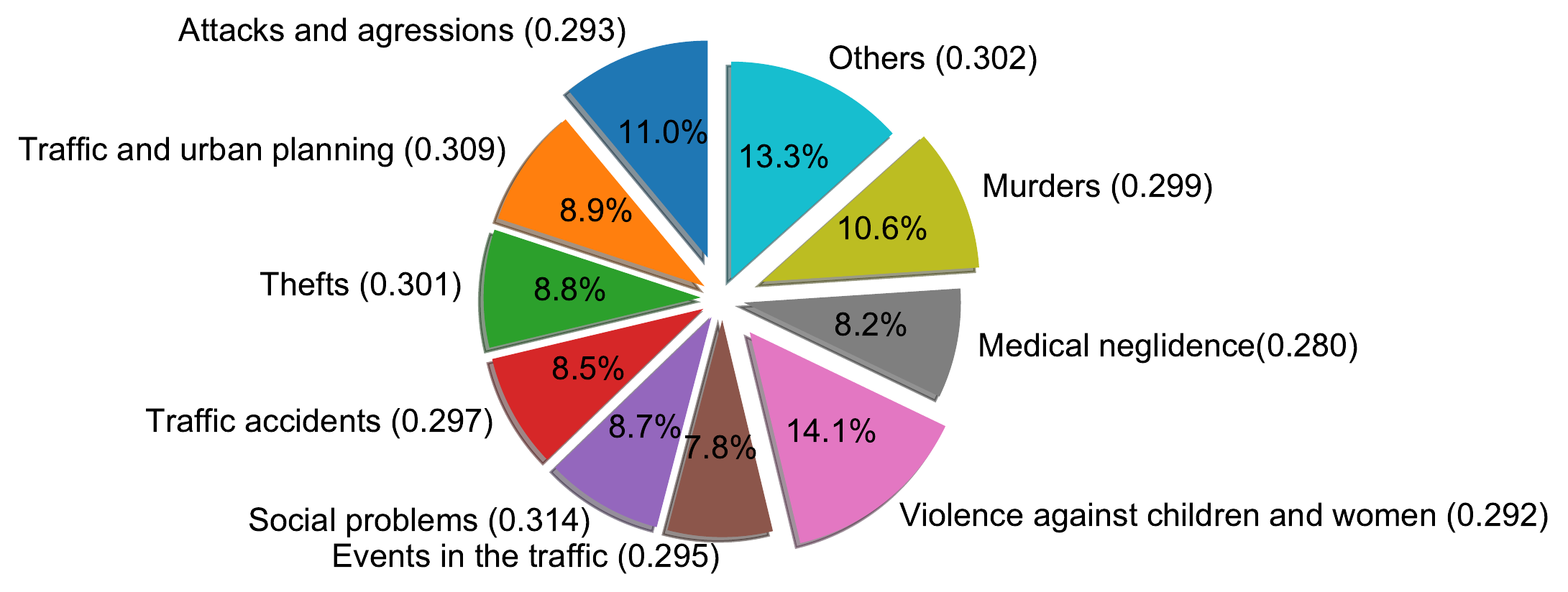}
    \caption{Same as~\cref{fig:pie_sixteen} but for 10 topics.}
    \label{fig:pie_ten}
\end{figure}

Finally, the distribution of the tweets along the 10 different topics are shown in \cref{fig:pie_ten}. We found in this number of topics, that the clusters correspond to more general topics, allowing a better separation of the tweets. It is noteworthy to state that the ``others'' cluster mixes different security problems such as aggression against animals, government infringements to provide services to people, among others.

\section{Conclusions\label{sec:conclusions}}

In this paper we presented the application of a latent topic discovery method at a fine-grained level to segment Colombian news published through Twitter in different sub-topics regarding security, crime and violence. We were able to find interpretable groups of tweets published by news-media giant @NoticiasRCN, where each group referred to different sorts of security, crime and violence issues.

We identified clear labels that summarise the content of the tweets belonging to each topic: attacks and aggressions, traffic and urban planning, thefts, traffic accidents, social problems, events in the traffic, violence against children and women, medical negligence, murders and others. An important application of the methodology presented in this paper is to detect violent events that go unreported to the police. Furthermore, the tools that were shown configure a critical channel for monitoring violent actions that attempt against the security of women, children, minorities and crime victims. Moreover, our method allows the automatic classification of new security-related tweets.

Our method contributes to the segmentation of tweets to better address issues in each security front. What we will develop in future work is the characterisation of people's reaction to different types of security, crime and violence related issues, and to identify violent behaviour in social networks, as this is a cornerstone to understanding social and cultural dynamics in our communities.

 \bibliographystyle{splncs04}
 \bibliography {mybibliography}
\end{document}